\documentstyle[preprint,prb,aps]{revtex}

\begin{document}

\draft
\title {RANDOM MATRIX THEORY APPROACH TO THE INTENSITY
DISTRIBUTIONS OF WAVES PROPAGATING IN A RANDOM MEDIUM}

\author{Eugene Kogan and Moshe Kaveh }

\address{ Jack and Pearl Resnick Institute of Advanced Technology,\\
Department of Physics, Bar-Ilan University, Ramat-Gan 52900, Israel}

\date{\today }

\maketitle
\begin{abstract}
Statistical properties of coherent radiation propagating in a quasi - 1D
random media is studied in the framework of random matrix theory. Distribution
functions for the total transmission coefficient and the angular
transmission coefficient are obtained.
\end{abstract}


The discovery of universal conductance fluctuations ( UCF ) \cite{r1,r2} has
induced a partial shift of the main interest in the studies of electronic
properties from the averaged values of physical quantities to their variance
and then to the whole distribution functions (see Altshuler {\it et
al.}\cite{r4}
and
references therein). Later it was demonstrated that UCF exist also for the
propagation of classical waves (e.\ g.\ light) through disordered systems.
\cite
{r3} In contrast to electronic measurements which can measure only the
conductance of a system, light experiments have the advantage of being able
to measure the angular and the total transmission coefficients for an
experimental
realization.

In our previous publication \cite{r9} we analyzed the problem of statistics
of radiation using diagrammatic techniques.It was rigorously
shown that the distribution function can be represented through the
contribution of connected diagrams only. This representation allowedto
develop a perturbation theory; in the frameworks of this theory it was
found that only for moderate values of the angular transmission coefficient
the distribution function
is a simple exponential, as predicted by Rayleigh statistics.
For larger values of intensity, the distribution function
differs drastically from a simple exponentialand it's asymptotical
behaviour is a stretched exponential decay. Also for the total transmission
coefficient the Gaussian distribution function was obtained.

An important step was made by Nieuwenhuizen and van Rossum. \cite{r14}
While in Ref.\ \onlinecite{r9}the perturbation series was truncated after
the second term, Nieuwenhuizen and van Rossum using diagrammatic techniques
combined with random matrix theory
managed to sum up the whole perturbation series, obtaining in particular
different stretched exponent for the angular transmission coefficient
distribution function and deviations from
the simple Gaussianfor the total transmission coefficient distribution
function.

In this paper we reproduce the results of Ref.\ \onlinecite{r14}
in the framework of the random matrix theory.
The approach is based on the analysis of the transfer matrix $R$
(see Stone {\it et al.} \cite{r10} and references therein).
Under the restrictions of flux conservation and time-reversal invariance,
this matrix can be represented in the form:
\begin{equation}
\label{x1}
R=\left(\matrix{ u & 0 \cr
0 & u^{\star}\cr}\right)
\left(\matrix{\sqrt{1+\lambda } & \sqrt{\lambda } \cr
\sqrt{\lambda } & \sqrt{1+\lambda }\cr }\right)
\left(\matrix{v & 0 \cr
0 & v^{\star} \cr} \right),
\end{equation}
where $u$ and $v$ are arbitrary $N\times N$ unitary matrices and $\lambda $ is
a
real, diagonal matrix with N positive elements $\lambda _1,\dots ,\lambda _N$,
where $N=W^2k^2$ is the number of transverse channels ($W^2$ is the area
of the sample). The $N\times N$ transmission matrix is given by
\begin{equation}
\label{x2}t=u \;\tau^{1/2}\;v,
\end{equation}
where $\tau \equiv \left( 1+\lambda \right) ^{-1}$.

In the isotropic approximation \cite{r11} an ensemble of $R$
 matrices is described by the
differential probability $dP(R)=P \left(\left\{ \tau \right\}\right)
\prod\limits_a d\tau_a d\mu (u)d\mu (v)$, where $d\mu (u)$ ($d\mu (v)$) is the
invariant measure of the unitary group U(N). This isotropic approximation is
rather strong assumption implying the perfect mode mixing but for a quasi-1D
systems it is
known to be good. \cite{r10}

The angular transmission coefficient $T_{ab}$,defined as the
ratio of the energy carried away by the transmitted wave with the transverse
wave vector $\vec q_b$ to the energy of the incident wave with the
transverse wave vector $\vec q_a$,
is given by $|t_{ab}|^2$, $t_{ab}$
being the $ab$ matrix element of Eq.\ (\ref{x2}). The n-th moment of $T_{ab}$
can be written down in the following way:
\begin{eqnarray}
\label{x4}
\langle T_{ab}^n \rangle & = &
 \sum\limits_{\{\alpha\},\{\beta\}}\langle(u_{a\alpha _1}\ldots
u_{a\alpha _n})(u_{a\beta _1}\ldots u_{a\beta _n})^{\star}
\rangle_0\langle(v_{b\alpha _1}\ldots v_{b\alpha _n})
(v_{b\beta _1}\ldots v_{b\beta _n})^{\star}\rangle_0 \nonumber \\
&& \times \langle(\tau _{\alpha _1}\ldots \tau_{\alpha _n}
\tau _{\beta _1}\ldots \tau _{\beta _n})^{1/2}\rangle_{\tau};
\end{eqnarray}
where the average indicated
by the index 0 is performed with the invariant measure of the unitary group and
$\langle X \rangle_{\tau}\equiv \int d\left\{ \tau \right\} \;P\left(
\left\{ \tau \right\} \right)\; X $.

It is known that to leading order in $1/N$ both real and imaginary components
 of $u_{a\alpha}$ and $v_{\beta b}$ are independently distributed Gaussian
variables with zero mean and variance $1/2N$. \cite{r15,r16} Then we can write
down
correlator $\langle(v_{b\alpha _1}\ldots
v_{b\alpha _n})(v_{b\beta _1}\ldots v_{b\beta _n})^{\star}\rangle_0$ as
the product ofcorrelators
$\langle v_{b\alpha}v_{b\beta}^{\star}\rangle_0 = \delta_{\alpha\beta}/N$
summed up with respect to all $n!$ possible pairings between $\{\alpha\}$
and $\{\beta\}$. So from Eq.\ (\ref{x4}) we get
\begin{equation}
\label{x5}
\langle T_{ab}^n\rangle = \frac {n!} {N^{n}}\sum_{\{\alpha\}}\langle
|u_{a\alpha
_1}|^2\ldots |u_{a\alpha _n}|^2\rangle_0
\langle \tau_{\alpha _1}\ldots \tau _{\alpha _n}\rangle_{\tau}
= \frac {n!} {N^{n}} \langle T_{a}^n\rangle,
\end{equation}
where
\begin{equation}
\label{x3a}T_a=\sum_{\alpha} |u_{a \alpha}|^2 \tau _{\alpha}.
\end{equation}
It can be easily seen, that $T_a$ is just the
total transmission coefficient: $T_a= \sum_b T_{ab}$.
In fact the n-th moment of the total transmission coefficient $\sum_b T_{ab}$
is
\begin{eqnarray}
\label{x49}
\langle \left(\sum\limits_b T_{ab}\right)^n\rangle & = &
\sum\limits_{\{\alpha\},\{\beta\}, \{b\}}
\langle (u_{a\alpha _1}\ldots
u_{a\alpha _n})(u_{a\beta _1}\ldots u_{a\beta _n})^{\star}\rangle_0
\langle(v_{b_1\alpha _1}\ldots v_{b_n\alpha _n})(v_{b_1\beta _1}
\ldots v_{b_n\beta _n})^{\star}\rangle_0 \nonumber \\
&& \times \langle(\tau _{\alpha _1}\ldots \tau_{\alpha _n}
\tau _{\beta _1}\ldots \tau _{\beta _n})^{1/2}\rangle_{\tau}.
\end{eqnarray}
To leading order in $1/N$
\begin{equation}
\label{x50}
\sum\limits_{\{b\}}\langle (v_{b_1\alpha _1}\ldots v_{b_n\alpha _n})
(v_{b_1\beta _1}\ldots v_{b_n\beta _n})^{\star}\rangle_0 =
\delta_{\alpha_1\beta_1}\ldots \delta_{\alpha_n\beta_n}
\end{equation}
(because the b-indexes are different we should take into account only one
pairing), and the right hand part of Eq.\ (\ref{x49})
is exactly $\langle T_{a}^n\rangle$.

Returning to Eq.\ (\ref{x3a}) we see that the distribution function $P(T_{a})$
can be written as an integration over eigenvalues and eigenvectors:
\begin{equation}
\label{x9a}
P(T_a)=\int d\tau _1\ldots \int d\tau _n\int dU\,\;\;P\left( \left\{
\tau \right\} \right)\;
\delta \left( T_a -\sum_{\alpha}|u_{a\alpha}|^2\tau _{\alpha}\right).
\end{equation}

It is convenient to work with the Laplace transform (we also measure $T_{a}$
in units of $\langle T_{a}\rangle = g/N$, where g is classical conductance):
\begin{equation}
\label{x12}
P \left(T_a \right)=
\int_{-i\infty }^{i\infty }\frac{ds}{2\pi i}\exp \left( sT_a\right)
\;F\,(s/g),
\end{equation}

Then easilycarrying out the integration with respect to $d U$, we find:
\begin{equation}
\label{x13}\;F(s)=\langle \;\prod_{\alpha =1}^N\frac 1{1+s\tau _{\alpha}}\;
\rangle_{\tau}.
\end{equation}

We are going to use an approximation of uniform distribution of the "charges"
$\nu_{\alpha}$, which are defined by the relation:
$\tau _{\alpha}=1/\cosh {}^2(\nu _{\alpha}/2)$.\cite{r10}
 That is, knowing that the distribution of
"charges" is statistically homogeneous,\cite{r10} instead of averaging with
respect to all possible configurations of "charges" we take into account only
one configuration - crystal lattice, which leads to the following relation:
\cite{r14}

\begin{equation}
\label{x18}\;\sum_{\alpha =1}^N f(\tau_{\alpha}) = g \int_0^{1}\frac{d\tau}
{2\tau\sqrt{1-\tau}}f(\tau).
\end{equation}
for any $f(\tau)$ which goes to zero when $\tau$ goes to zero.
Then fromEq.\ (\ref{x13}) we get
\begin{equation}
\label{x14}\;F(s)=\exp \left[ -g\int_0^1\frac{d\tau}{2\tau\sqrt{1-\tau}}
\ln \left(1+ s\tau\right) \right] =
\exp \left[ -g\ln {}^2\left( \sqrt{1+s}+\sqrt{s}\right) \right],
\end{equation}
which exactly coincides with the result of
 Ref.\ \onlinecite{r14}.
Eq.\ (\ref{x14})
gives in particular Gaussian behavior for $T_a \approx 1$:
\begin{equation}
P(T_a)\approx \sqrt{\frac{3g}{4\pi} } \exp[ -\frac{3g}{4} (T_a-1)^2]
\end{equation}
and simple exponential decay for large $T_a$:
\begin{equation}
P(T_a)\sim \exp(-gT_a).
\end{equation}

Now let us return to Eq.\ (\ref{x5}). As it is known \cite{r9} it means
\begin{equation}
\label{x8}
P(T_{ab})=\int_0^\infty d T_a \;P(T_a )~{\frac
1{T_a} }\exp \left( -{\ \frac{T_{ab}}{T_a} }\right)
\end{equation}
(we measure $T_{ab}$
in units of $\langle T_{ab}\rangle = g/N^2$, where g is classical conductance).
This distribution functioncan be described as the Rayleigh distribution
function for the angular transmission coefficient
but with some effective averaged value which inturn fluctuates
around the real averaged value and the latter fluctuations are described by the
total transmission coefficient distribution function. \cite{r9,r14}
Eq.\ (\ref{x8}) gives in particular Rayleigh statistics
\begin{equation}
\label{x32}
P(T_{ab}) \approx \exp \left(-T_{ab}\right).
\end{equation}
for $T_{ab} \ll \sqrt{g}$,~~
\cite {r9} and stretched exponential tail
\begin{equation}
\label{x33}
P(T_{ab}) \sim \exp \left(-2 \sqrt{g T_{ab}}\right).
\end{equation}
for $T_{ab} \gg g$. \cite{r14}

Having in mind the comparison of the
theoretical result with an experiment it is convenient to express $g$
through the
first two moments either of the totalor of the
angular transmission coefficient distribution function.
Calculating the coefficient
 before $s^2$ in the expansion of the exponent in the right hand part of the
 Eq.\ (\ref{x14}) we get:
\begin{equation}
\label{x21}\; \frac{<T_{a}^2>}{<T_{a}>^2}-1 = \frac{2}{3g}\;,
\end{equation}
and
\begin{equation}
\label{x20}
\frac{<T_{ab}^2>}{<T_{ab}>^2}-2 = \frac{4}{3g}\;.
\end{equation}
which exactly coincides with the result of Ref.\ \onlinecite{r11}.

In conclusion we want to discuss the
difference between the statistics of total transmission coefficient and
statistics of conductance $g=\sum_m \tau_m$.
Taking into account the bimodal distribution of $\tau$ we may say,
at least qualitatively, that the conductance is simply the number
of "open"channels \cite{r12,r17}: $g = N_{eff}$ . The total
 transmission coefficient is also the sum with respect to "open" channels but
each
channel comes with a random weight.
So the Gaussian law for the total transmission
coefficient distribution function is just
 the manifestation of the Central Limit Theorem, which is true when
$N_{eff} \rightarrow \infty$.
In the paper we are taking into account the finitness of theparameter
$N_{eff}$, which is important
in particular for obtaining correct asimptotics.
 On the other hand, the conductance fluctuations are determined by
the strongly suppressed fluctuations of the number of open channels, which in
our case can be neglected. This principal difference between the two statistics
wouldalso manifest itself if one tries to go beyond quasi-1D.
 While the eigenvalue distribution
 (and hence the conductance distribution function)
 can be not very sensitive to
the dimensionality and stay bimodal as long as we are in a diffusive regime,
 \cite{r18}
the isotropic approximation which was essential
in obtainingEq. (\ref{x14}) ceases to be valid beyond quasi-1D. \cite{r19}

\section*{Acknowledgements}
Discussions with Y. Avishai, M. C. W. van Rossum and B. Shapiro
are gratefully acknowledged.
 The authors also acknowledge the financial support of the
Israeli Academy of Sciences and of Schottenstein Center.

\end{document}